\documentclass[11pt, a4paper]{article}

\usepackage[english]{babel}
\usepackage[utf8]{inputenc}
\usepackage{jcappub_ver}
\usepackage{enumerate}
\usepackage{amsbsy}
\usepackage{amsmath} 
\usepackage{graphics}
\usepackage{mathrsfs}
\usepackage{wrapfig}
\usepackage{tabularx}
\usepackage{calc}

\providecommand{\f}[2]{\frac{{#1}}{{#2}}}

\newcommand{\bea}{\begin{eqnarray}} \newcommand{\eea}{\end{eqnarray}}
\newcommand{\el}{\nonumber \\}
\newcommand{\re}[1]{(\ref{#1})}

\newcommand{\pat}{\partial}

\renewcommand{\sec}[1]{section \ref{#1}}
\newcommand{\fig}[1]{figure \ref{#1}}
\newcommand{\tab}[1]{table \ref{#1}}
\newcommand{\brt}[1]{[#1]}
\newcommand{\para}{\paragraph}

\renewcommand{\a}{\alpha}
\renewcommand{\b}{\beta}
\renewcommand{\c}{\gamma}
\renewcommand{\d}{\delta}

\renewcommand{\l}{\lambda}

\newcommand{\rmd}{\mathrm{d}}

\newcommand{\nonum}{\\}
\newcommand{\etal} {et al.\ }
\newcommand{\ie}{i.e.\ }

\newcommand{\bx}{\boldsymbol{x}}
\newcommand{\bk}{\boldsymbol{k}}

\newcommand{\sP}{{\cal P}}
\newcommand{\sH}{{\cal H}}

\newcommand{\dphi}{\d{\vphi}}
\newcommand{\phih}{\d\hat{\vphi}}
\newcommand{\phib}{\bar{\vphi}}

\newcommand{\dphigi}{\dphi_{\mathrm{GI}}}

\newcommand{\ndot}{\dot{n}}

\newcommand{\av}[1]{\langle{#1}\rangle}
\newcommand{\bra}[1]{\langle{#1}|}
\newcommand{\ket}[1]{|{#1}\rangle}
\newcommand{\vac}{\ket{0}}

\newcommand{\cR}{\mathcal{R}}
\newcommand{\sR}{{^{(3)}R}}

\newcommand{\PRD}[1]{{\it Phys. Rev.} {\bf D#1}}

\newcommand{\PRL}[1]{{\it Phys. Rev. Lett.} {\bf #1}}

\newcommand{\NPB}[1]{{\it Nucl. Phys.} {\bf B#1}}

\newcommand{\PLB}[1]{{\it Phys. Lett.} {\bf B#1}}
\newcommand{\MNRAS}[1]{{\it Mon. Not. Roy. Astron. Soc.} {\bf #1}}
\newcommand{\APJ}[1]{{\it Astrophys. J.} {\bf #1}}

\newcommand{\JHEP}[1]{{\it JHEP} {\bf #1}}
\newcommand{\CQG}[1]{{\it Class. Quant. Grav.} {\bf #1}}
\newcommand{\GRG}[1]{{\it Gen. Rel. Grav.} {\bf #1}}

\newcommand{\PROG}[1]{{\it Prog. Theor. Phys.} {\bf #1}}

\newcommand{\IJMPD}[1]{{\it Int. J. Mod. Phys.} {\bf D#1}}

\newcommand{\ham}{\mathcal{H}}
\newcommand{\epsilonh}{{\epsilon_H}}
\newcommand{\etah}{{\eta_H}}
\newcommand{\vphi}{\varphi}
\newcommand{\hfphi}{\ham \dfrac{\delta \vphi_{\bk}}{\vphi'}}					
\newcommand{\fphid}{\dfrac{\delta \vphi'_{\bk}}{\vphi'}}

\title{Inflation without quantum gravity}

\author{Tommi Markkanen,}

\emailAdd{tommi {\it dot} markkanen {\it at} helsinki {\it dot} fi}

\author{Syksy R\"{a}s\"{a}nen}

\emailAdd{syksy {\it dot} rasanen {\it at} iki {\it dot} fi}

\author{and Pyry Wahlman}

\emailAdd{pyry {\it dot} wahlman {\it at} helsinki {\it dot} fi}

\affiliation{University of Helsinki, Department of Physics \\
and Helsinki Institute of Physics \\
P.O. Box 64, FIN-00014 University of Helsinki, Finland}

\abstract{
It is sometimes argued that observation of tensor modes from inflation
would provide the first evidence for quantum gravity.
However, in the usual inflationary formalism, also the
scalar modes involve quantised metric perturbations.
We consider the issue in a semiclassical setup
in which only matter is quantised, and spacetime is classical. 
We assume that the state collapses on a spacelike hypersurface,
and find that the spectrum of scalar perturbations depends on the
hypersurface. For reasonable choices, we can
recover the usual inflationary predictions for scalar
perturbations in minimally coupled single-field models.
In models where non-minimal coupling to gravity is
important and the field value is sub-Planckian, we
do not get a nearly scale-invariant spectrum of scalar perturbations.
As gravitational waves are only produced at second order, the
tensor-to-scalar ratio is negligible.
We conclude that detection of inflationary gravitational waves
would indeed be needed to have observational evidence of quantisation of gravity.
}

\begin{document}

\maketitle
  
\setcounter{tocdepth}{2}

\setcounter{secnumdepth}{3}

\section{Introduction} \label{sec:intro}

\para{Inflation and the quantisation of gravity.}

The recent observation by the BICEP2 experiment \cite{BICEP2}
of B-mode polarisation of the cosmic microwave background in excess
of the signal due to lensing \cite{lensing}, at first credited to
gravitational waves generated during inflation, turned out to be
due to Galactic foregrounds \cite{fore, joint}.
(There are also other possible cosmological sources of B-modes,
such as magnetic fields \cite{Bonvin:2014}, topological defects
\cite{defect} and self-ordering scalar fields \cite{Durrer:2014},
though these could not have explained the signal.)

The claimed detection of inflationary tensor modes was hailed as
the first observational confirmation of quantum gravity,
because inflationary tensor perturbations are generated from
quantum fluctuations of the metric\footnote{Note that observation
of primordial gravitational waves with an amplitude of $10^{-5}$
would not, in itself, prove quantisation of gravity, unlike
argued in \cite{Krauss}.
First order scalar perturbations generate tensor perturbations at second
order, so if the amplitude of scalar perturbations were $10^{-3}$ to $10^{-2}$,
gravitational waves with an amplitude of $10^{-5}$ would be
generated by this classical process. Arguments based on dimensional
analysis and the amplitude of primordial gravitational waves alone
do not provide evidence for quantum gravity.}.
Since inflationary gravitational waves have not been observed, does
this mean that there is no observational evidence for quantum gravity?
In the usual treatment of inflation, this is not the case,
because scalar perturbations of the metric are also quantised.
More precisely, the relevant scalar quantity is the
Sasaki-Mukhanov variable \cite{SM}, which is the linear combination of
matter and metric perturbations that satisfies canonical commutation
relations. (Neither the gauge-invariant scalar metric perturbation
nor the gauge-invariant scalar field perturbation satisfy canonical
commutation relations on their own; see e.g. \cite{Eltzner:2013}.)
The resulting predictions for the spectrum of scalar modes are in
excellent agreement with observations.
However, if it is possible to obtain the same results
by quantising only the matter variables, then gravitational waves
would really be needed to establish quantisation of the metric.
The issue is complicated by the fact that we do not know which
approximation to quantum gravity is valid during inflation.

\para{Semiclassical quantum gravity.}

Inflation is the only area of physics where it has been possible
to observationally probe the interface between general relativity
and quantum theory \cite{Starobinsky:1980, inflation}.
Indeed, it seems that an indeterministic theory is required to
explain the origin of inhomogeneity and anisotropy in the universe.
Deterministically it is only possible to refer anisotropy and
inhomogeneity to an earlier inhomogeneous and anisotropic state,
absent constraints that only allow a unique initial state
or dynamical laws with preferred locations and directions.
The indeterminism of quantum mechanics makes it possible to
break the symmetry of the initial state by collapse, in which only one
member of the statistically homogeneous and isotropic
initial distribution of possibilities is realised.
This connects the generation of inflationary perturbations to
state collapse in the primordial universe.

There is no complete theory of quantum gravity, only
different approximate or speculative formulations.
Quantisation of metric perturbations in the Einstein-Hilbert action
(with added higher order curvature counterterms) around a classical
background leads to a non-renormalisable theory.
However, it is possible for perturbatively non-renormalisable
theories to be non-perturbatively well-defined \cite{Gawedzki}.
In the case of gravity, one possibility is asymptotic
safety, which means that the renormalisation group flow has a fixed point
corresponding to a finite-dimensional hypersurface in the
space of coupling constants \cite{Weinberg:2009, safety}.
Even non-renormalisable theories may be treated at low
energies using effective field theory \cite{Weinberg:2009, Gomis:1995}
and perturbative quantum gravity is no exception \cite{Donoghue:2012}.
Quantum corrections can be organised in a series of higher order
curvature terms, possibly along with important infrared effects
not captured by such an expansion \cite{Tsamis, Mazur:2001}.

Quantising perturbations of the metric around a classical background,
as done in the usual treatment of inflation, is often called the
semiclassical approximation.
However, the term semiclassical is also used to refer to a distinct
approach to quantum field theory on curved spacetime, in which
quantum matter evolves in a classical spacetime \cite{Parker:2009}.
This formalism has been developed to great sophistication in
terms of algebraic quantum field theory \cite{algebra}.
In this approach, quantum fields are coupled to
the classical metric via the expectation value of the
energy-momentum tensor being proportional to the Einstein tensor.
We use this theory to study whether it is possible to
reproduce the successes of the usual inflationary formalism
without quantising the metric.
If it turns out that the usual inflationary predictions
for the scalar modes can be reproduced without quantising metric
perturbations, then observation of inflationary gravitational waves
would (in the inflationary context) be necessary in order
to conclude that gravity is quantised.
(We are here neglecting the issue of loop corrections due to gravitons,
which are below present observational sensitivity.)

This is not expected to be a valid approximation close to the Planck
scale or when the variance is large and typical realisations are
far from the mean (see also \cite{Pinamonti:2010}).
Another issue is that the expectation value of the energy-momentum
tensor, and thus the metric, changes discontinuously when the state
collapses, which can lead to problems.
For example, if state collapse can be precipitated locally, causality
can be violated, as this makes it possible to send signals
with infinite speed. (If the metric is also in a superposition state
until collapse, there is no such problem.)
We will only consider a single collapse event, and do not specify the
collapse mechanism. Also, in inflation the amplitude of typical perturbations
is small, so different realisations are not far from each other.
Whether or not the domain of validity of the semiclassical approximation
extends to inflation can only be settled by a more complete
theory of quantum gravity or by direct comparison to observation.
Inflation without quantised metric perturbations has been previously
considered in \cite{Sudarskycl}.

We take the distribution of perturbations in the initial state to be
homogeneous and isotropic. The expectation value of the energy-momentum
tensor is then spatially homogeneous and isotropic, and the spacetime
is taken to be the homogeneous and isotropic
Friedmann--Robertson--Walker (FRW) universe.
Once the state collapses into a specific realisation of the quantum
mechanical distribution, the expectation value of the energy-momentum
tensor is no longer homogeneous and isotropic, and will source metric 
perturbations. We treat the system
classically after the collapse, so the quantum mechanical
distribution of realisations becomes a classical
distribution in space, as usual for inflation.
Unlike in the usual treatment, the resulting power spectrum
of the curvature perturbation depends on the details of the collapse.
As usual in quantum mechanics, we assume that the state collapses on
a spacelike hypersurface, and we investigate different choices of hypersurface.
(See \cite{Sudarskycl, Sudarskyqm, Martin:2012} for scenarios where
the collapse happens at different times for different wavemodes.)

In \sec{sec:form} we give the action and the equations of motion
in the semiclassical formalism.
In \sec{sec:match} we consider matching across the hypersurface
of collapse and calculate the spectrum for different choices.
We show that for minimally coupled single field inflation models,
we can recover the usual predictions for scalar perturbations,
up to slow-roll suppressed corrections.
If non-minimal coupling to gravity is important,
we do not get a nearly scale-invariant spectrum
(at least for sub-Planckian field values).
We summarise our results in \sec{sec:con}.

\section{Semiclassical inflation} \label{sec:form}

\subsection{Action and equations of motion} \label{sec:act}

\para{Classical case.}
We consider scalar field matter that may be non-minimally coupled to gravity,
\bea \label{action}
  S = \int\rmd^4 x \sqrt{-g} \left( \frac{M^2 + \xi \vphi^2}{2} R - \frac{1}{2} g^{\a\b} \pat_\a\varphi \pat_\b\varphi - V(\vphi) \right) \ ,
\eea

\noindent where $M$ and $\xi$ are constants and $R$ is the Ricci scalar.
We use the metric signature $(-+++)$, the conventions
of \cite{Misner:1973} for the curvature functions, and the sign
convention where $\xi=-\frac{1}{6}$ is the conformally coupled case.
If both matter and spacetime were classical, we would
obtain the equations of motion by varying the action \re{action}
with respect to the metric and the scalar field.

If the metric is taken to be the only independent gravity variable,
variation of \re{action} with respect to the metric gives
\bea \label{Einstein}
  G_{\a\b} = \frac{1}{M^2} T_{\a\b} \ ,
\eea

\noindent where $G_{\a\b}$ is the Einstein tensor and $T_{\a\b}$ is
the energy-momentum tensor, and the energy-momentum tensor is
\bea \label{Tc1}
  T_{\a\b} = S_{\a\b} - \xi ( G_{\a\b} - \nabla_\a \nabla_\b + g_{\a\b} \Box ) \vphi^2 \ ,
\eea

\noindent where
$S_{\a\b}\equiv  \pat_\a\vphi \pat_\b\vphi - g_{\a\b} [ \frac{1}{2} g^{\c\d} \pat_\c\vphi \pat_\d\vphi + V(\vphi) ]$.
In the Palatini formulation, where the metric and the connection are
independent variables, the energy-momentum tensor is different from
\re{Tc1}, unless the non-minimal coupling $\xi$ vanishes. We come back
to this difference when we consider Higgs inflation in \sec{sec:model}.
The energy-momentum tensor \re{Tc1} is quadratic in the field,
and thus well suited to quantisation. For cosmological analysis, it
is more convenient to replace $G_{\a\b}$ using \re{Einstein} to obtain
\cite{Madsen:1988}
\bea \label{Tc2}
  T_{\a\b} = \frac{ M^2 }{ M^2 + \xi \vphi^2 } \left[  S_{\a\b} + \xi ( \nabla_\a \nabla_\b - g_{\a\b} \Box ) \vphi^2 \right] \ ,
\eea

\noindent The effective Planck mass corresponds
to $\sqrt{M^2 + \xi \vphi^2}$, so for $|\xi|\vphi^2\ll M^2$,
it is close to $M$. We adopt units such that $M=1$.

Variation of \re{action} with respect to the scalar field gives
\bea \label{eom}
  \Box \vphi + \xi R \vphi - \frac{\rmd V}{\rmd\vphi} = 0 \ .
\eea

\para{Semiclassical case.}

We consider quantum matter in a classical spacetime.
The relation between the two is given by the semiclassical
Einstein equation
\bea \label{Einsteinq}
  G_{\a\b} = \av{T_{\a\b}} \ ,
\eea

\noindent where $\av{T_{\a\b}}$ is the expectation value of the
energy-momentum tensor. We neglect higher order curvature terms,
which arise when renormalising a quantum field in curved
spacetime. For small curvature, their effect is small,
apart from the feature that they change the order of the differential
equations and typically destabilise the solutions \cite{stability}.
However, instabilities related to higher derivatives are
presumably beyond the range of validity of our semiclassical
approach \cite{Simon}\footnote{See \cite{Gorka:2012} for an
example of a class of theories with infinitely high derivatives
that have a well-defined initial value problem.}.

We assume that the expectation value of the energy-momentum tensor
before the collapse is homogeneous and isotropic and the spacetime
is FRW. We do not consider how inflation has started and how
the spacetime has become homogeneous and isotropic
\cite{Goldwirth:1992, Trodden, Ellis}.
This inverts the usual relation between the symmetry of the background
space and the symmetry of the quantum mechanical distribution.
Usually the distribution inherits homogeneity and isotropy
from the background space, whereas in our case spacetime is FRW because
of the symmetry of the distribution\footnote{In the first inflationary
model, the initial state was actually assumed to be homogeneous
and isotropic \cite{Starobinsky:1980}.}.
We assume that inflation has lasted sufficiently long that spatial curvature
can be neglected, so the metric is
\bea \label{metricFRW}
  \rmd s^2 = a(\eta)^2 \left[ - \rmd \eta^2 + \d_{ij} \rmd x^i \rmd x^j \right] \ .
\eea

\noindent We denote $\sH\equiv a'/a$, where prime denotes derivative with
respect to conformal time $\eta$.

We write the field operator in terms of a homogeneous
expectation value and perturbation as $\hat{\varphi}=\phib+\phih$,
with $\langle\hat{\varphi}\rangle=\phib$.
We neglect quantum corrections to the equations of motion,
which are small during inflation at least in chaotic inflation
models \cite{qcorr}, which we consider in \sec{sec:model}.
Expanding the action \re{action} around $\phib$ to quadratic order
in perturbations gives
\bea \label{actionp}
  S=&\int\rmd ^4x\sqrt{-g}\left( \frac{1 + \xi\phib^2}{2} R - \frac{1}{2} g^{\a\b} \pat_\a\phib \pat_\b\phib - V(\phib) \right)\nonumber \\ &-\f{1}{2}\int d ^4x\sqrt{-g} \phih\left(-\Box-\xi R+ \frac{\rmd^2 V(\phib)}{\rmd\phib^2}\right)\phih \ .
\eea

\noindent The field perturbation is written in terms of annihilation and
creation operators $\hat a_{\bk}$ and $\hat a^\dagger_{\bk}$
(respectively) as usual,
\bea \label{phiop}
  \phih(\eta, \bx) = \sum_{\bk} \left( \hat a_{\bk} u_{\bk}(\eta, \bx) + \hat a^\dagger_{\bk} u_{\bk}^*(\eta, \bx) \right) \ .
\eea

\noindent (Strictly speaking, \re{phiop} is not a well-defined operator
and has to be smeared over a spacetime patch to obtain a well-defined
quantity \cite{algebra}.)
We assume that the spacetime allows a decomposition in terms of $\bk$-modes.
As we restrict our attention to FRW spacetime, this is not a problem,
and we have (with some abuse of notation)
$u_{\bk}(\eta, \bx)=u_{\bk}(\eta) e^{i\bk\cdot\bx}$.

>From \re{actionp} we get the equations of motion for
$\phib$ and $u_{\bk}$ by varying with respect to $\phib$ and $\phih$,
\bea
  \label{eomphib} \phib'' + 2 \sH \phib' - 6 \xi ( \sH' + \sH^2 ) \phib + a^2 \frac{\rmd V(\phib)}{\rmd\phib} &=& 0 \\
  \label{eomu} u_{\bk}'' + 2 \sH u_{\bk}' - 6 \xi ( \sH' + \sH^2 ) u_{\bk} + a^2 \frac{\rmd^2 V(\phib)}{\rmd\phib^2} u_{\bk} &=& 0 \ ,
\eea

\noindent where in \re{eomu} we only consider modes
with super-Hubble wavelengths. The normalisation of $u_{\bk}$
is fixed by the fact that $\phih$ satisfies canonical commutation
relations\footnote{As mentioned in \sec{sec:intro}, in the usual
treatment of inflation, the gauge-invariant field perturbation does
not satisfy canonical commutation relations, nor does the
gauge-invariant scalar metric perturbation. As the two are related by a constraint, it would be impossible
for both of them to satisfy canonical commutation relations; see e.g.
\cite{Eltzner:2013}.}.
A crucial feature of the semi-classical mode equation \re{eomu} is that during inflation for a minimally coupled theory it coincides with the fully quantised result up to slow-roll corrections\footnote{This feature that the semiclassical and fully quantum treatments agree up to small corrections is not generic: For example in the usual treatment of the hydrogen atom one must invoke Gauss's law as an operator identity and not as a relation between two expectation values in order to obtain correct results.}.

An important question is in which state the expectation value
is taken. We follow the usual assumption that the system is in
the Bunch-Davies vacuum, which is a Hadamard state.
As the Hamiltonian does not commute with the field operator, this
vacuum is not an eigenstate of the field operator, so the field
does not have a well-defined value. This is the origin of
inflationary quantum fluctuations.

Instead of expanding the action around a non-zero
vacuum expectation value, we could assume that the expectation
value of the field in the Bunch-Davies vacuum $\vac$ is zero,
$\hat\vphi=\phih$.
In this case, we could obtain a non-vanishing expectation value
by taking the system to be in a coherent state instead.
Specifically, we can consider the state
$\hat U\vac\equiv e^{N \hat a_{\bf 0} + N^* \hat a_{\bf 0}^\dagger}\vac$,
where $\hat a_{\bf 0}$ is the annihilation operator corresponding
to the zero mode and $N(\eta)$ is a function chosen so that
$\av{\hat\varphi}=\av{\phih}=2 \mathrm{Re}(N u_{\bf 0})$
satisfies \re{eom}, so we have $\langle\hat{\varphi}\rangle=\phib$, as before.
(In other words, the time-dependence of $N$ is adjusted by hand to
correct the evolution of the zero mode to follow the full non-linear
equation of motion \re{eomphib} instead of its linearisation \re{eomu}.)
We can go back to the situation with a non-zero vacuum expectation value
by transforming the field instead of transforming the state
(\ie switching from the Schr\"{o}dinger representation to the Heisenberg
representation), which gives
$\phih\rightarrow \hat U^\dagger \phih \hat U = 2 \mathrm{Re}(N u_{\bf 0}) + \phih = \phib + \phih$.
However, the situation with the coherent state is not physically
fully equivalent to the case where we start from the action
\re{actionp}, because the time evolution of the coherent state
function $N$ is not determined by the classical equations of motion.

Either way, we have
\bea
  \av{\hat\vphi^n} &=& \sum_{m=0}^{[n/2]} \frac{n!}{(2 m)! (n-2m)!} \phib^{n-2m} \bra{0} \phih^{2m} \vac \el
  &=& \sum_{m=0}^{[n/2]} \frac{n!}{2 m! (n-2m)!} \phib^{n-2m} \left( \int_0^\infty\frac{\rmd k}{k} \sP_{\phih}(k) \right)^m \ ,
\eea

\noindent where $[n/2]$ is the integer part of $n/2$.
On the second line we have assumed that the statistics of the quantum
fluctuations are Gaussian, homogeneous and isotropic and written
$\bra{0}\phih^2\vac=\int_0^\infty\frac{\rmd k}{k} \sP_{\phih}(k)$,
where $\sP_{\phih}$ is the power spectrum of the vacuum
fluctuations of $\phih$.
If the typical amplitude of the power spectrum is small
and it decays at both small and large $k$, so that
$\int_0^\infty\frac{\rmd k}{k} \sP_{\phih}(k)$ is small,
we have $\bra{0}\phih^2\vac\ll\phib^2$. If the time derivatives of the mode
functions are not much larger than the time derivatives of $\phib$,
the expectation value of the energy-momentum tensor in \re{Einsteinq}
reduces to the classical form \re{Tc1} in terms of $\phib$
(Strictly speaking, the situation is more complex, and the
argument requires renormalising divergent quantities, and depends
on the form of the inflaton potential \cite{qcorr}\footnote{From
(\ref{actionp}) we get the quantum correction to the energy-momentum
tensor as (using cosmic time instead of
conformal time; dot denotes derivative with respect to cosmic time)\begin{align}
\langle\hat{T}^Q_{00}\rangle &=\sum_{\bk}\bigg\{\f{1}{2}\bigg[\vert\dot{v}_\mathbf{k}\vert^2+\bigg( {\frac{\mathbf{k}^2}{a^2}+\frac{\rmd^2 V(\phib)}{\rmd\phib^2} }\bigg)\vert v_\mathbf{k}\vert^2\bigg]-\xi\bigg[G_{00}+3\frac{\dot{a}}{a}\partial_0\bigg]\vert v_\mathbf{k}\vert^2\bigg\}\nonumber \\&=\sum_{\mathbf{q}}\left\{b_1(\mathbf{q})+ b_2(\mathbf{q}) G_{00}+b_3(\mathbf{q})\frac{\rmd^2 V(\phib)}{\rmd\phib^2}+\mathcal{O}(\dot{v}_\mathbf{q})\right\}\, ,\label{eq:TQ}
\end{align}
where $\mathbf{q}\equiv\mathbf{k}/a$. The coefficients $b(\mathbf{q})$ depend on $v_\mathbf{q}$, and we have used the normalized mode functions ${v}_\mathbf{k}\equiv{u}_\mathbf{k}a^{-3/2}$ in de Sitter space \cite{Parker:2009}. When $\dot{u}_\mathbf{q}$ is small, the $b(\mathbf{q})$'s give (divergent) constants that can be absorbed in the bare action, provided that $V(\phib)$ is a polynomial in $\phib$. The first two terms in the second line of (\ref{eq:TQ}) are absorbed in the cosmological constant and Newton's constant, respectively, and the rest contribute to the matter action.}.)

\subsection{From homogeneity and isotropy to perturbations} \label{sec:pert}

\para{Metric and equations of motion.}

Before the collapse, the field is in a superposition state
of different field values,
and its probability distribution is homogeneous and isotropic.
When the state collapses, the field takes an almost definite value
at each point in space, and homogeneity and isotropy are broken.
(If the field value were exactly definite, its time derivative
would be completely indefinite due to the uncertainty principle.
For classical-looking states, the probability distribution of
both the field and its time derivative is highly peaked, but not
exactly definite.)
The expectation value of the energy-momentum tensor jumps
discontinuously to having perturbations, which source
perturbations in the metric.
Scalar field matter does not source vector and tensor
perturbations at first order, so we only have scalar
perturbations, and the metric can be written as
\bea \label{metricpert}
  \rmd s^2 = a(\eta)^2 \left[ - ( 1 + 2 A ) \rmd \eta^2 + 2 B,_i \rmd \eta \rmd x^i + \left( \d_{ij} - 2 \d_{ij} \psi + 2 E,_{ij} \right) \rmd x^i \rmd x^j \right] \ ,
\eea

\noindent and we denote $C\equiv B - E'$.
After the collapse, we treat the expectation value of
the quantum field as a classical quantity as per usual, and we
split it into background plus perturbation,
$\av{\hat\vphi}\equiv\vphi=\phib(\eta)+\dphi(\eta,\bx)$.
(Note the slight difference of notation from the pre-collapse era:
in both cases, $\phib$ denotes a homogeneous background field,
but after the collapse, the expectation value of the field
has perturbations around the mean.)
The background evolution is given by \re{eomphib} together
with the Einstein equation \re{Einstein} for the background,
\bea \label{EinsteinFRW}
  3 \sH^2 &=& \frac{1}{1+\xi\phib^2} \left( \frac{1}{2} \phib'^2 + a^2 V - 6 \xi \sH \phib \phib' \right) \ .
\eea

For the perturbations, the Einstein equation
\re{Einstein} and the field equation of motion \re{eom}
written in terms of Fourier modes reduce to (we only
consider super-Hubble modes)
\bea
  \label{eomdphi} && \dphi_{\bk}'' + 2 \sH \dphi_{\bk}' - 6 \xi ( \sH' + \sH^2 ) \dphi_{\bk} + a^2 \frac{\rmd^2 V}{\rmd\phib^2} \dphi_{\bk} 
  = - 2 a^2 \frac{\rmd V}{\rmd\phib} A_{\bk} + \phib' ( A_{\bk}' + 3 \psi_{\bk}' - k^2 C_{\bk} ) \el
  && - \xi \phib \left( 6 \psi_{\bk}'' + 18 \sH \psi_{\bk}' + 6 \sH A_{\bk}' + k^2 [ 4 \psi_{\bk} - 2 A_{\bk} - 2 C_{\bk}' - 6 \sH C_{\bk} ] \right) \\
  \label{psi} && \psi_{\bk} = A_{\bk} + C_{\bk}' + 2 \sH C_{\bk} + 2 \frac{\xi \phib}{1 + \xi \phib^2} ( \dphi_{\bk} + \phib' C_{\bk} ) \\
  \label{psi'} && \psi_{\bk}' + \sH A_{\bk} = \frac{1}{ 1 + \xi \phib^2 }  \left( \frac{1}{2} \phib' \dphi_{\bk} + \xi [ \phib' \dphi_{\bk} - \sH \phib \dphi_{\bk} + \phib \dphi' - \phib \phib' A_{\bk} ] \right) \\
  \label{C} && \sH C_{\bk} = \frac{1 + \xi\phib^2}{1 + \xi \phib^2 + \xi \sH^{-1} \phib \phib'} \psi_{\bk} \el
  && + \frac{1}{2} \frac{ 1 +  \xi \phib^2 + 6 \xi^2 \phib^2}{( 1 + \xi \phib^2 ) ( 1 + \xi \phib^2 + \xi \sH^{-1} \phib \phib')} \left( \frac{\phib'}{k} \right)^2 \left( [ \eta_H - \epsilon_H ] \sH \frac{\dphi_{\bk}}{\phib'} + \frac{\dphi_{\bk}'}{\phib'} - A_{\bk} \right)
\ ,
\eea

\noindent where we have defined
$\epsilon_H\equiv - (\sH' - \sH^2)/\sH^2$,
$\eta_H\equiv 1 - \phib''/(\sH\phib') + \epsilon_H$.
We also denote
$\eta_{H2}\equiv 2 - \phib'''/(\sH^2\phib') + 2 \epsilon_H - 3 \eta_H$.\footnote{In terms of cosmic time $t$, $\epsilon_H=-\pat_t H/H^2$,
$\eta_H-\epsilon_H=-\pat_t^2\phib/(H\pat_t\phib)$
and $\eta_{H2}=-\pat_t^3\phib/(H^2\pat_t\phib)$,
where $H\equiv\sH/a$.}
We have not assumed that these quantities are small.
In slow-roll (and for minimal coupling)
they reduce to the usual slow-roll parameters $\epsilon$,
$\eta$ and $\xi_2-4\epsilon^2-\eta^2+\epsilon\eta$,
respectively.
The comoving curvature perturbation on large scales is
(for proof of conservation in the case $\xi=0$, see
e.g. \cite{Gordon:2001}, section 2.2)
\bea \label{R}
  \cR = - \psi - \sH \frac{[ ( 1 + 2 \xi ) \phib' - 2 \xi \sH \phib ] \dphi + 2 \xi \phib \dphi' - 2 \xi \phib \phib' A}{( 1 + 2 \xi ) \phib'^2 - 2 \xi \sH \phib \phib' ( 1 - \epsilon_H + \eta_H )} \ .
\eea

\noindent In the case $\xi=0$, this reduces to the usual result
$\cR = - \psi - \sH \dphi/\phib' $.

\section{Matching across the collapse} \label{sec:match}

\subsection{Hypersurface of collapse} \label{sec:hyper}

\para{Squeezing, decoherence and collapse.}

During inflation quantum modes are stretched to super-Hubble
wavelengths and become squeezed as the ratio of the constant
and decaying solutions of the mode equation grows exponentially
\cite{squeezing}. The modes decohere and the quantum mechanical
distribution takes the shape of a classical stochastic
distribution \cite{decoherence}.
However, it is not understood how an approximately classical,
definite-seeming universe emerges, \ie how the state collapses,
though presumably this happens after squeezing and decoherence.
Some suggested collapse theories have been applied to inflation,
but the matter remains unsettled \cite{Sudarskycl, Sudarskyqm, Martin:2012}.
Note that the issue of decoherence (suppression of quantum mechanical
interference terms) is different from that of collapse (definite outcome).

In the usual formulation of inflation where both metric and matter
perturbations are quantised, the quantum mechanical distribution
of the Sasaki-Mukhanov variable (or the comoving curvature perturbation
$\cR$, treated as a quantum variable) is simply equated with its classical
distribution, without considering the collapse process.
We instead equate the pre-collapse quantum mechanical power spectrum
of the field with the post-collapse power spectrum of the
classical field, $\sP_{\hat\vphi}=\sP_{\vphi}$ or, equivalently,
$\sP_{\phih}=\sP_{\dphi}$, on some spacelike hypersurface of collapse.
The field is discontinuous at collapse, but its power spectrum
is continuous, and the power spectrum of its time derivative
can also be taken to be continuous, so we can match $\sP_{\phih'}=\sP_{\dphi'}$.
Discontinuity of the metric perturbations implies, via \re{eomdphi},
that the power spectra of second and higher order time derivatives
of the field are discontinuous across the collapse.

Unlike in \cite{Sudarskycl, Sudarskyqm, Martin:2012},
collapse conditions are not defined for individual modes
in momentum space, but instead all wavelengths collapse
simultaneously. Squeezing depends on the wavelength
of the mode, so sub-Hubble modes will not have become squeezed
and decohered at the time of collapse, and their distribution could be
very different from the classical stochastic case.
Wavelengths of such modes are extremely small compared to
present-day cosmological scales, as inflation typically lasts for
a few dozen e-folds after the observable modes become super-Hubble.
It is not clear whether their non-classical distribution
would have any observable consequences. Possible signals from very small
scales include primordial black holes \cite{Green:2014} and
gravitational waves.

In contrast to the usual inflationary treatment, the
power spectrum of the comoving curvature perturbation $\cR$
depends on which hypersurface the collapse happens.
In the pre-collapse era, space is exactly homogeneous and isotropic,
and the slice can be defined in terms of Killing vectors,
but after the collapse there is no exact symmetry, and the division
into an FRW background slice plus perturbations is simply a
matter of convenience \cite{gauge}.
On the collapse hypersurface, the unique exactly homogeneous
and isotropic pre-collapse spatial slice is matched onto
one of the infinite number of possible background post-collapse
spatial slices.
Equivalently, we have to specify on which hypersurface
the classical variable $\dphi$ is defined.
This ambiguity is related to the fact that $\dphi$ is not
gauge-invariant. However, the choice of the hypersurface of collapse
is not a gauge choice, and different hypersurfaces lead to different
power spectra for $\cR$.
Viewed physically, we have to specify which physical quantity
is constant on the hypersurface of collapse.

\para{Choice of hypersurface.}

In the absence of a description of the collapse process, we
simply consider different spatial hypersurfaces of collapse
and see how the power spectrum of the comoving curvature perturbation
depends on the choice.
In principle, it is possible to get any desired power spectrum
by choosing the appropriate $\psi$ on the hypersurface of collapse.
Nevertheless, some choices of hypersurface seem more natural than others.
In this subsection, we put $\xi=0$, because we will see in \sec{sec:model}
that the kind of hypersurfaces we consider do not lead to a nearly
scale-invariant spectrum if the non-minimal coupling is important,
except possibly in the case $|\xi|\ll1$, $|\phib|\gg1$.
We specify a hypersurface by setting conditions on the metric
or on some physical quantities. From those conditions
we solve for $\psi$ in terms of $\dphi$ and $\dphi'$ using
\re{eomdphi}--\re{C}, and use \re{R} to obtain $\cR$ in terms of
$\dphi$ and $\dphi'$. There are two unknowns, $A$ and $C$.
If the hypersurface condition fixes $C$ (and does not contain
inverse spatial derivatives of $\dphi$ and $\dphi'$), we obtain, in the
super-Hubble limit,
\bea \label{giveC}
  \cR_{\bk} = - ( 1 + \eta_H - \epsilon_H ) \sH \frac{\dphi_{\bk}}{\phib'} - \frac{\dphi'_{\bk}}{\phib'} - C_{\bk}' - 2 \sH C_{\bk} \ ,
\eea

\noindent where all quantities are evaluated at the time of collapse,
not at Hubble crossing. If the hypersurface condition fixes $A$ instead,
we obtain
\bea \label{giveA}
  \cR_{\bk} = \frac{1}{\epsilon_H} ( f_{\bk}' + 2 \sH f_{\bk} ) \ ,
\eea

\noindent where
$f_{\bk}\equiv \epsilon_H \sH k^{-2} \left( [ \eta_H - \epsilon_H ] \sH \frac{\dphi_{\bk}}{\phib'} + \frac{\dphi'_{\bk}}{\phib'} - A_{\bk} \right)$.
Again, all quantities are evaluated at the time of collapse.
Note that in this case $\cR$ will depend on $\dphi''$,
unless $A$ contains exactly one factor of $\frac{\dphi'}{\phib'}$.  
As the power spectrum of $\dphi''$ is discontinuous at collapse,
such a hypersurface will not yield a well-defined power spectrum.

We first consider the simple possibility of matching
$\sP_{\phih}=\sP_{\dphigi}$, where
$\dphigi\equiv\dphi+\phib' C$ is the gauge invariant field perturbation.
This corresponds to setting $C=0$, so if the collapse happens during
slow-roll, and perturbations evolve slowly,
$|\dphi'_{\bk}|\ll\sH|\dphi_{\bk}|$, \re{giveC}
shows that the power spectrum of $\cR$ in terms of $\dphi$ is the
same as in the usual inflationary case in the spatially flat gauge, $\psi=0$.
In the spatially flat gauge the field equation of motion
also reduces to \re{eomu} at zeroth order in slow-roll. We therefore
obtain the same power spectrum as in the usual case,
up to slow-roll suppressed corrections.

\para{Covariant quantities.}

\begin{table}

  \center
  \renewcommand{\arraystretch}{2.1}

  \begin{tabular}{|>{$}l<{$}|>{$}l<{$}|}
    \hline
    \text{Constant} & \text{Curvature perturbation $\cR_{\bk}$} \\
    \hline
    C = 0 & - \big( 1 - \chi_1 \big) \hfphi - \fphid \\
    \theta & - 3 \epsilonh \dfrac{\ham^2}{k^2} \left[ \big( 3 + \chi_1 \big) \hfphi + \fphid \right] \\
    \sigma = 0 & - 2 \big( 1 - \chi_1 \big) \hfphi - 2 \fphid \\
    \dot\vphi & - \dfrac{\ham^2}{k^2} \left[ \big( 3 \chi_1 +\chi_2 \big) \hfphi + \chi_1 \fphid \right] \\
    \rho & 3 \dfrac{\ham^2}{k^2} \left[ \big( 3 +\chi_1 \big) \hfphi + \fphid \right] \\
    p & - \dfrac{\ham^2}{k^2} \left[ \big( 9 +9 \chi_1 +2 \chi_2 \big) \hfphi + \big( 3 + 2 \chi_1 \big) \fphid \right] \\
    R & - 3 \dfrac{\ham^2}{k^2} \left[ \big( 6 + 5 \chi_1 + \chi_2 \big) \hfphi + \big( 2 + \chi_1 \big) \fphid \right] \\
    \sR = 0 & - 2 \hfphi \\[7pt]


  \hline

  \end{tabular}

\caption{Curvature perturbation for different choices of what is
kept constant (in the case of $C$, $\sigma$ and $\sR$, kept zero)
on the hypersurface of collapse.
All quantities are evaluated at the time of collapse.
Here we taken $\xi=0$ and denoted
$\chi_1\equiv\epsilonh-\etah$ and $\chi_2\equiv\epsilonh\chi_1-\eta_{H2}$.}

\label{tab:HS}

\end{table}

Let us now consider collapse hypersurfaces defined
in terms of covariant physical quantities.
Consider a frame-independent physical scalar quantity $S$.
The unit vector orthogonal to the hypersurface of constant $S$,
assumed to be spacelike, is
\bea \label{n}
  n^\a = \frac{1}{\sqrt{ - g^{\c\d} \pat_\c S \pat_\d S }} g^{\a\b} \pat_\b S \ .
\eea

It has been argued that in single-field inflation the preferred
time variable is the field \cite{Geshnizjani:2002}.
This would correspond to taking the collapse hypersurface to be the one
of constant $\vphi$. However, the condition $\dphi=0$ does not give a
well-defined spectrum for $\cR$ on the hypersurface of collapse (and if it did,
the result would be zero). We can instead keep constant some other quantity
defined in terms of $n^\a$ for the choice $S=\vphi$.
The gradient of $n^\a$ can be decomposed in terms of the expansion rate
$\theta$, shear $\sigma_{\a\b}$ and acceleration
$\ndot^\a$, where dot refers to derivative along $n^\a$ \cite{covariant}.
>From $\sigma_{\a\b}$ and $\ndot^\a$, we can form the scalar quantities
$\sigma\equiv\sqrt{\frac{1}{2}\sigma^{\a\b}\sigma_{\a\b}}$
and $\ndot\equiv\sqrt{\ndot^\a\ndot_\a}$.
Some simple choices of quantities to keep constant on the
hypersurface are $\dot\vphi$, constant expansion rate $\theta$,
shear $\sigma$, spatial curvature scalar $\sR$,
energy density $\rho$ or pressure $p$.
In the case of $\sigma$ and $\sR$, they are actually zero
on the hypersurface.
Taking the spacetime Ricci scalar $R$ to be constant 
defines a hypersurface in a frame-independent way
(\ie without having to specify $n^\a$).
Obviously, other choices are possible, but those that
involve $\dphi''$ or derivatives of metric perturbations,
which are not defined at the moment of collapse, will
not yield a power spectrum for $\cR$.
Examples of such ill-defined choices include the hypersurface of
constant volume acceleration $\dot\theta+\frac{1}{3}\theta^2$ or
constant (in fact, zero) acceleration $\ndot$ (taking $S=\vphi$ in both cases).

In \tab{tab:HS}, we give results for the above choices of
collapse hypersurface.
We have two kinds of modifications to the spectrum
compared to the usual inflationary case. The first possibility
is that the spectrum is multiplied by the factor $(k/\sH)^{-4}$,
evaluated at collapse.
Without inflation producing a very blue spectrum for $\dphi$,
with a spectral index close to $n=5$, the resulting spectrum for $\cR$
is not close to scale-invariant, in strong disagreement with observations.
Also, for observable modes we have $k\ll\sH$ at collapse,
and the amplitude of $\sP_{\cR}$ would be exponentially
larger than the amplitude of $\sP_{\dphi}$.
In addition to $\sP_{\phih}=\sP_{\dphigi}$,
this leaves the hypersurfaces of constant (in fact, vanishing)
$\sigma$ and $\sR$ as possibly viable choices.

The other modification is a uniform change in the amplitude.
The magnitude depends on the hypersurface of
collapse and on when the collapse occurs.
Presumably the state does not collapse before modes in the observable range
have become squeezed and decohered, and it may be that the
collapse does not happen until the decay of the inflaton field.
Conceivably, it could even happen later, but we only follow the system
up to the beginning of inflaton oscillations, and do not consider decay.
The change in the amplitude is determined by the values of
$\epsilon_H-\eta_H$, $\sH\dphi/\phib'$ and $\dphi'/\phib'$
at collapse, so it depends on the inflationary model.
We first look at two simple minimally coupled inflationary
models, and then consider the non-minimally coupled case.

\subsection{Inflation models} \label{sec:model}

\para{Minimal coupling.}

As examples, we consider the minimally coupled ($\xi=0$)
inflation models defined by the potentials
$V=\frac{1}{2}m^2\vphi^2$ and $V=\frac{1}{4}\l\vphi^4$.
The background equations \re{eomphib} and \re{EinsteinFRW} are the same
as in the usual inflationary formalism, but the mode equation \re{eomu}
is different from the usual equation \re{eomdphi} due to the absence of metric
perturbations.
However, as discussed above, in the usual inflationary case the contribution
of metric perturbations to \re{eomdphi} vanishes in the spatially flat
gauge ($\psi=0$) to zeroth order in slow-roll.
Therefore we get essentially the same results for $\sP_{\dphi}$
as in the usual case, if the state collapses during slow-roll.
If we match $\sP_{\phih}=\sP_{\dphigi}$
across the collapse, we also get the usual result for
$\sP_{\cR}$, up to slow-roll suppressed corrections.
For the hypersurface of constant $\sigma$ or $\sR$, the amplitude of
$\sP_{\cR}$ is increased by a factor of two compared to the usual case.

\begin{center}
\begin{figure}

\includegraphics[width=0.45\textwidth]{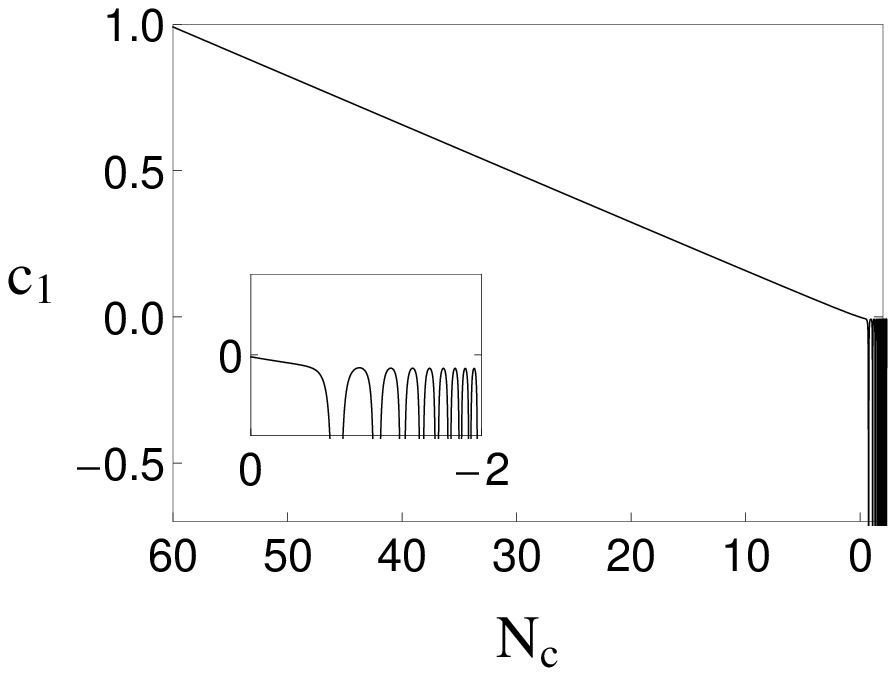}
\includegraphics[width=0.45\textwidth]{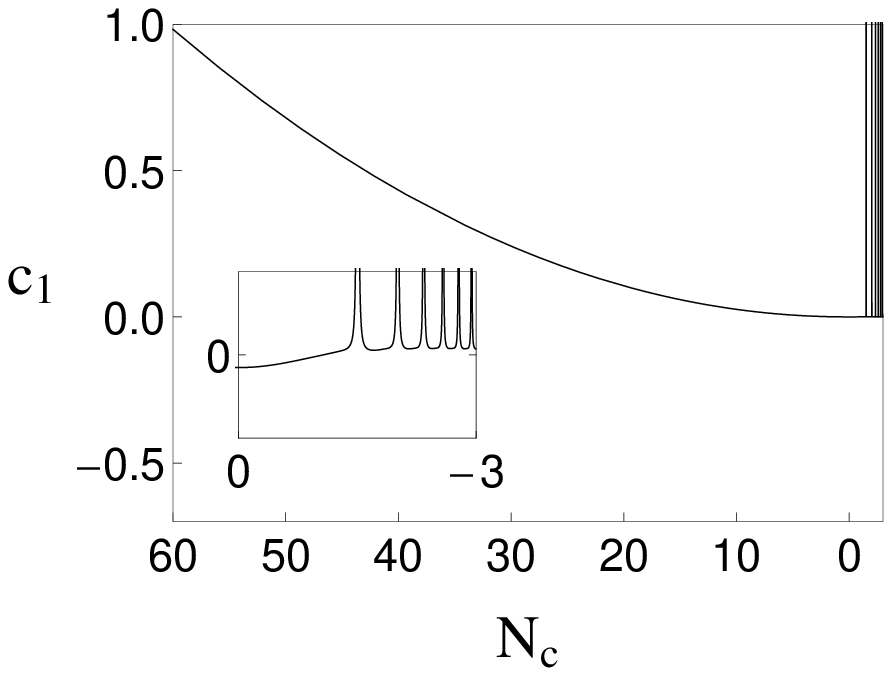}
\includegraphics[width=0.45\textwidth]{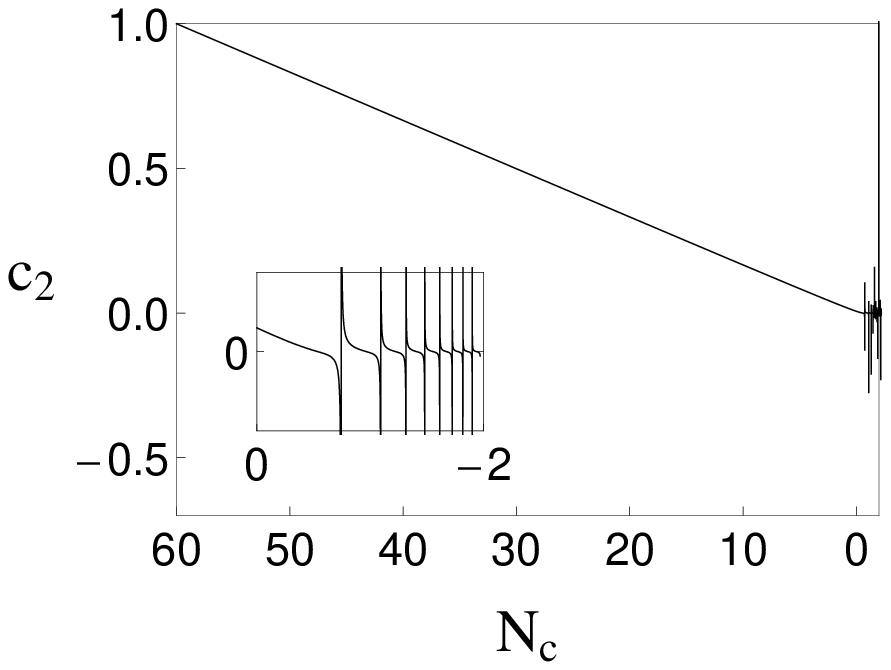}\hspace{1.35cm}
\includegraphics[width=0.45\textwidth]{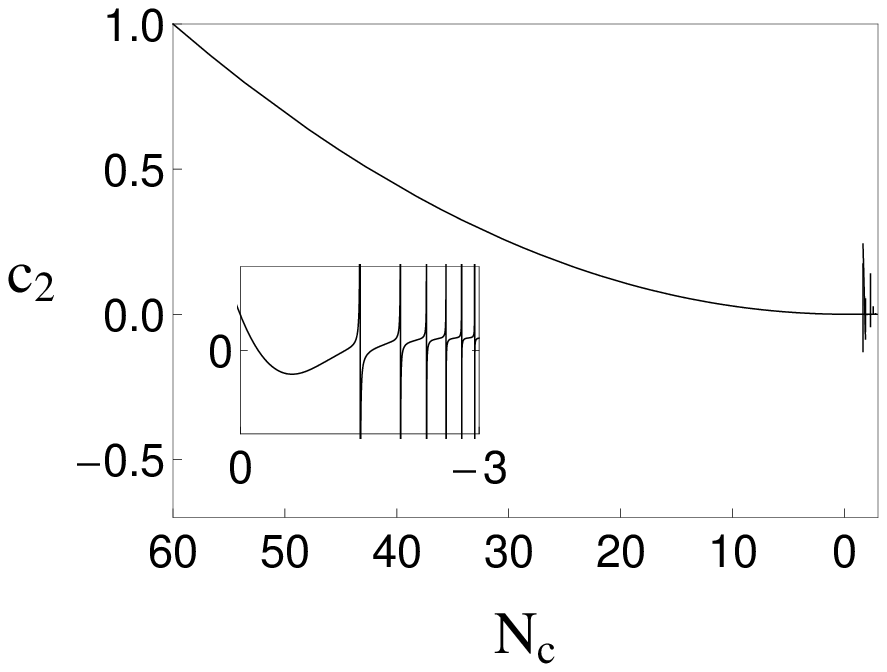}

\caption{The quantity $c_i\equiv-\cR/(\sH_{60}\dphi_{60}/\varphi_{60}')$
as a function $N_{\mathrm{c}}$, the number of e-folds until
the end of inflation at collapse; the subscript 60 refers
to the terms being evaluated at 60 e-folds until
the end of inflation.
The top shows $c_1$, which corresponds to collapse
on the hypersurface of constant $\dphigi$.
The bottom shows $c_2$, which corresponds to collapse
on the hypersurface of constant $\sR$.
The left and right sides correspond to the potentials
$V=\frac{1}{2}m^2\vphi^2$ and $V=\frac{1}{4}\l\vphi^4$,
respectively. Insets show evolution after the end
of inflation in detail.}

\label{fig:amp}

\end{figure}
\end{center}

If the state collapses when the field oscillates after inflation,
the absolute value of the amplitude can change up or down
by an arbitrary amount, for any of these three hypersurfaces.
(Obviously, the range of validity of our perturbative calculation
does not extend to non-perturbatively large amplitudes, so the most
we can say is that the amplitude can become non-perturbatively large.)
In \fig{fig:amp} we show $\cR$ as a function of $N_{\mathrm{c}}$, the
number of e-folds until the end of inflation at the moment of
collapse (so $N_{\mathrm{c}}=0$ corresponds to $\epsilon_H=1$),
normalised to the usual spatially flat gauge result $-\sH\dphi/\phib'$
evaluated at 60 e-foldings until the end of inflation.
At the top we show the result for $\sP_{\phih}=\sP_{\dphigi}$
(for the hypersurface of constant $\sigma$, the result is
multiplied by 2, as seen in \tab{tab:HS}).
If the collapse happens deep into inflation, the change
is small, but if the state collapses when the inflaton field is
oscillating, the amplitude can change up or down by any amount,
because $\phib'$ passes through zero. For the potential
$V=\frac{1}{2}m^2\vphi^2$, shown on the left side, $\cR$
first diverges at $N_{\mathrm{c}}=-0.7$, and then
undergoes divergent oscillations with a negative amplitude.
For $V=\frac{1}{4}\l\vphi^4$, the behaviour is similar, with
$\cR$ first diverging at $N_{\mathrm{c}}=-1.5$, and then undergoing
divergent oscillations with a positive amplitude.
At the bottom we show the result for collapse
on the hypersurface of constant $\sR$. The behaviour is similar,
with the amplitude undergoing divergent oscillations.

If the amplitude of $\cR$ relative to $\dphi$ is reduced,
$m$ or $\l$ could be larger than in the normal inflationary setup.
However, non-Gaussianity due to second order field perturbations
would be correspondingly larger, so the change in the amplitude
is restricted to be smaller than one order of magnitude \cite{Planck:NG}.
If the amplitude of $\cR$ is instead enhanced, $\dphi$ could
be much smaller then usual, and non-Gaussianity from this source
would be suppressed from its usual amplitude (which is of order unity).
This can in principle be tested observationally, given sufficiently
precise observations and control over other sources of non-Gaussianity.

\para{Non-minimal coupling.}

Let us now consider the case when non-minimal coupling is
important. Eliminating $\rmd^2 V/\rmd\phib^2$ in the
mode equation \re{eomu} by using \re{eomphib}, we have
\bea \label{eomuxi}
  && \!\!\!\!\!\!\!\! \frac{\rmd^2 u_{\bk}}{\rmd N^2} + 3 \frac{\rmd u_{\bk}}{\rmd N} + \left( 3 \eta_H + \eta_{H2}  \right) u_{\bk} \el
  && \!\!\!\!\!\!\!\! + 6 \frac{ \xi \phib^2 }{ 1 + \xi \phib^2 } \left( \xi [ 4 - 2 \epsilon_H + 3 \eta_H - \eta_{H2} ] - \frac{\phib'}{\sH \phib} [ 2 + 3 \xi + ( 1 + 3\xi ) ( \epsilon_H - \eta_H ) ] \right) u_{\bk} = 0 \ ,
\eea

\noindent where $N\equiv\ln a$. The modes evolve slowly if the absolute
value of the dimensionless effective mass term (the factors
in front of $u_{\bk}$) is much smaller than unity.
If the background evolution is such that the slow-roll parameters
can be neglected, then smallness of the effective mass requires either that
\mbox{$24 \xi^2 \phib^2/(1 + \xi \phib^2)\ll1$},
or that the factor in parenthesis on the second line of \re{eomuxi} is small.
Considering the first possibility, if $|\xi|\phib^2\ll1$,
we also have $\xi^2\phib^2\ll1$, and the non-minimal coupling
has negligible effect on the dynamics.
If instead $|\xi|\phib^2\gtrsim1$ and $|\xi|\ll1$, we have $|\phib|\gg1$.
Vanishing of the factor in parenthesis also leads to the same condition.
Thus, if the non-minimal coupling plays a role in inflation,
the modes do not evolve slowly, at least for sub-Planckian field values, 
so the field perturbation spectrum is not close to scale-invariant.
According to \re{psi}--\re{C} and \re{R}, the resulting
spectrum of curvature perturbations is not close scale-invariant either.
(Matching on a hypersurface that changes the spectrum by the factor
$(k/\sH)^{-4}$ would not compensate to make the spectrum of $\cR$
scale-invariant without tuning.)
In particular, this rules out inflation with the Standard Model Higgs
field \cite{nonminimal, SMHiggs}.
In the super-Planckian case with $|\xi|\ll1$, the amplitude of
primordial perturbations is enhanced relative to the minimally coupled
case (if the non-minimal coupling has any role), so the potential has to
be even flatter than usual, in addition to other possible problems such as
Planck-scale suppressed higher order terms.

Considering the Palatini formalism, in which the metric and the
connection are independent variables, would change the equations of motion
\cite{Palatini}, but this would likely not affect the conclusion.
In the slow-roll limit, the effective mass term is only
modified by terms that are either quadratic or linear in $\phib'$.
Its smallness would again give an independent constraint
on the evolution of the background, which is in general not compatible
with the background evolution equations. This is in contrast to the
minimally coupled inflationary case, in which slow-roll is a sufficient
condition for small effective mass.

The reason that inflation driven by non-minimal coupling to gravity
does not produce a scale-invariant spectrum is that quantised
perturbations of the Ricci scalar $R$ are crucial in the mode
equation of motion. In the case when both field and metric
perturbations are quantised, slow-roll for the background implies
that the effective mass of the perturbations modes is small.
In our case, when only the field is quantised, the background Ricci
scalar contributes to the background equations, but its perturbations
do not contribute to the mode equation,
so slow-roll does not guarantee small effective mass.
We conclude that in Higgs inflation, a nearly scale-invariant
spectrum of scalar perturbations requires quantisation of metric
perturbations, unlike in minimally coupled inflation.

\section{Conclusions} \label{sec:con}

\para{Inflation in a classical spacetime and the role of collapse.}

It has been argued that observation of gravitational waves from
inflation would provide the first evidence for quantum gravity, because
inflationary generation of tensor modes with amplitude close to that
of the scalar modes requires quantising metric perturbations.
In the usual formalism of inflation, also the scalar modes
involve quantised metric perturbations, so gravitational waves
do not provide qualitatively new evidence in this regard. 
However, we do not know which approximation of quantum gravity
is valid during inflation, and if it is possible to reproduce the
predictions of the usual inflationary formalism without
quantising the metric, then gravitational waves would be needed
to establish quantisation of gravity.

We have considered inflation in a semiclassical gravity, where
the matter is quantised and the metric is classical.
We have assumed that the state collapses on a spacelike hypersurface
for all wavelengths, in contrast to some previous work where the collapse time
depended on the wavelength \cite{Sudarskycl, Sudarskyqm, Martin:2012}.
This means that small wavelength modes collapse before they
become squeezed and decohere; it is not clear whether this has
observational consequences. Whereas the spectrum for the field
perturbation before the collapse is unambiguous,
the inherited spectrum of the comoving curvature perturbation depends
on the hypersurface of collapse.
Selecting the correct hypersurface by some physical principle remains an open question, our main goal was to study whether this setup is tenable in principle.
We have considered some simple possibilities and
found that for minimally coupled single field inflation models it is
possible to recover the usual inflationary results up to leading
order in slow-roll parameters, provided the collapse happens during inflation.
If the state collapses during preheating, the amplitude of the power spectrum
can change by an arbitrary amount. This would change the amplitude of
non-Gaussianity due to second order inflaton field perturbations,
and there are tight constraints on possible increase of the amplitude,
though it can be decreased without limit.

In inflationary models where non-minimal coupling to gravity plays
an important role, the situation is different, at least when the field
amplitude is sub-Planckian.
In that case, metric perturbations play a crucial role in the equation
of motion of the mode functions, and it is not possible to generate
a nearly scale-invariant spectrum of scalar perturbations without
quantising the metric.
Thus, models like Higgs inflation require quantum gravity.

In summary, without a detection of gravitational waves or confirmation
that non-minimal coupling to gravity plays an important role in inflaton
dynamics, successful inflation does not require quantising any part
of the metric.
Therefore, without the detection of inflationary gravitational waves,
we cannot conclude (from tree-level results in inflation) that gravity is quantised.

\acknowledgments

SR thanks Shaun Hotchkiss for disagreements.
TM is supported by the Mikael Bj\"{o}rnberg Memorial Fund
and the Academy of Finland through grant 1134018.
PW is supported by the Finnish Cultural Foundation.

\end{document}